\documentclass[prb,twocolumn]{revtex4}
\usepackage{times}
\usepackage{bbm}
\usepackage[dvips]{graphicx}
\usepackage{latexsym,amsmath,amssymb,bm,euscript}
\usepackage[dvips]{color}
\usepackage{multirow}

\def\ra{\rangle} 
\def\la{\langle} 

\begin{document}

\title{Phases and phase transitions in a $U(1)\times U(1)$ system with $\theta=2\pi/3$ mutual statistics}
\date{\today}
\pacs{}

\author{Scott D. Geraedts}
\author{Olexei I. Motrunich}
\affiliation{Department of Physics, California Institute of Technology, Pasadena, California 91125, USA}

\begin{abstract}
We study a $U(1)\times U(1)$ system with short-range interactions and mutual $\theta=2\pi/3$ statistics in (2+1) dimensions. We are able to reformulate the model to eliminate the sign problem, and perform a Monte Carlo study. We find a phase diagram containing a phase with only small loops and two phases with one species of proliferated loop. We also find a phase where both species of loop condense, but without any gapless modes. Lastly, when the energy cost of loops becomes small we find a phase which is a condensate of bound states, each made up of three particles of one species and a vortex of the other. We define several exact reformulations of the model, which allow us to precisely describe each phase in terms of gapped excitations. We propose field-theoretic descriptions of the phases and phase transitions, which are particularly interesting on the ``self-dual'' line where both species have identical interactions. We also define irreducible responses useful for describing the phases.
\end{abstract}
\maketitle

\section{Introduction}
One of the hallmarks of topological quantum phases is that they have anyonic excitations, which can be viewed as particles with statistical interactions.  Examples include quasiparticles in the Fractional Quantum Hall Effect\cite{Stern2008}, spinon and vison excitations in $Z_2$ spin liquids,\cite{Read1989, Wen91, Kitaev2003, SenthilFisher_Z2} and excitations in a variety of interesting fractionalized systems.\cite{Levin2005, Nayak2008_rmp, Z3frac, LevinStern2009}  It is also fruitful to ask about possible new phases that such particles can have, as a way to access proximate phases and phase transitions involving topological quantum states.\cite{Zhang1989, LeeFisher1989, Tupitsyn2010, Barkeshli2010, Kou2009, Wen2000, Burnell2011, Gils2009}

Unfortunately, direct Monte Carlo studies are hampered by the sign problem.  It turns of that some such systems allow reformulations where they become free of the sign problem and can be studied using unbiased numerical approaches.  Interesting questions include, for example, what phases can result if there are two species of particles with mutual statistics that are both trying to condense.
In this work, we pursue such a study of the effects of a statistical interaction on a model of two species of integer-valued loops with short-range interactions. We are able to reformulate this model so that it can be studied on a lattice using Monte Carlo techniques. Previously,\cite{Geraedts} we studied a model with two species of loops and mutual $\pi$ statistics, which is also of interest in effective field theories of frustrated antiferromagnets\cite{Senthil2006_theta, deccp_science, deccp_prb, Xu2009, KamalMurthy, shortlight} and other areas.\cite{Hansson2004, Kou2008, Cho2011, Xu2011}  We would like to extend this to study systems with general statistical angle $\theta$. We have found that $\theta=\pi$ is a special case, and the properties of such models are qualitatively different when $\theta\neq\pi$. In this work we study $\theta=2\pi/3$, and the results should exhibit behavior similar to that for general $\theta\neq\pi$.

\begin{figure}[t]
\input{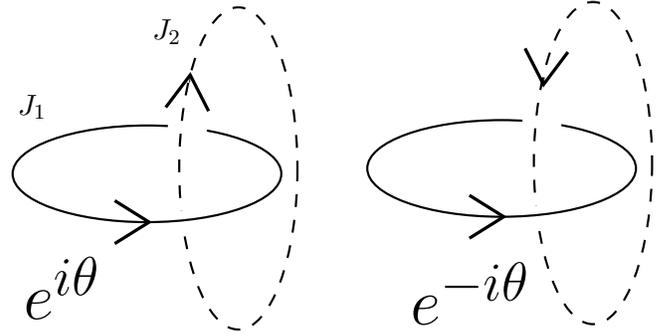}
\caption{The contribution to the partition function is multiplied by a phase $e^{i\theta}$ for each cross-linking of the two currents shown in the figure on the left. If we change the direction of one of the currents and get the figure on the right, the phase is $e^{-i\theta}$. When considering symmetries of our model for $\theta\neq\pi$, we must only consider operations that leave the relative orientation of the current loops unchanged. }
\label{loops}
\end{figure}

Our model can be precisely described by the following action:
\begin{equation}
S[\vec{J}_1,\vec{J}_2] = \sum_r \frac{\vec{J}_1(r)^2}{2 t_1}
+ \sum_R \frac{\vec{J}_2(R)^2}{2 t_2}
+ i \theta \sum_r \vec{J}_1(r) \cdot \vec{p}_2(r) ~.
\label{action}
\end{equation}
The index $r$ refers to sites on a cubic lattice (the ``direct'' lattice), and $R$ refers to sites on another, inter-penetrating cubic lattice (the ``dual'' lattice).\cite{Fradkin_SL2Z, Kantor1991, footnoteone}
$J_{1\mu}(r)$ is an integer-valued current on a link $r,r+\hat{\mu}$ of the direct cubic lattice, $J_{2\mu}(R)$ is integer-valued current on a link $R,R+\hat{\mu}$ of the dual cubic lattice. We use schematic vector notation so that $\vec{J}_1$ and $\vec{J}_2$ represent these conserved integer-valued currents, and $\vec{\nabla} \cdot \vec{J}_1 = 0, \vec{\nabla} \cdot \vec{J}_2 = 0$. In the partition sum, a given current configuration obtains a phase factor $e^{i\theta}$ or $e^{-i\theta}$ for each cross-linking of the two loop systems, dependent on the relative orientation of the current loops, as shown in Fig. \ref{loops}. This is realized in the last term of Eq.~(\ref{action}), by including an auxiliary ``gauge field'' $\vec{p}_2$, defined on the direct lattice, whose flux encodes the $\vec{J}_2$ currents, $\vec{J}_2 = \vec{\nabla} \times \vec{p}_2$.  

\begin{figure}[t]
\includegraphics[angle=-90,width=\linewidth]{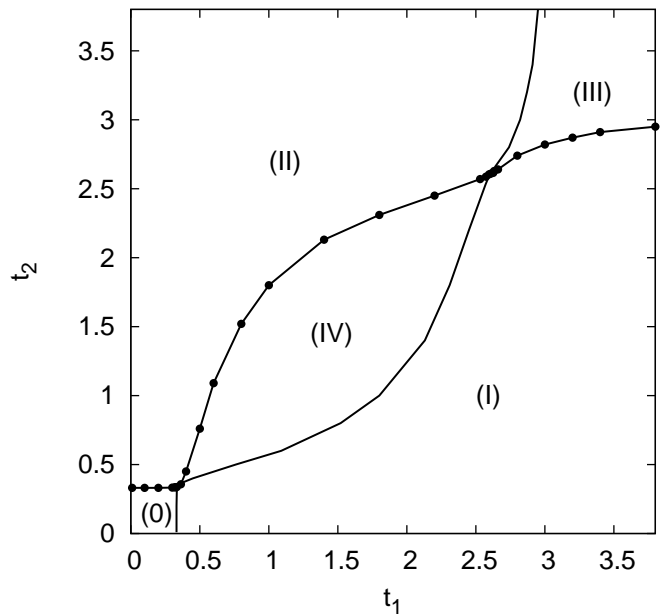}
\caption{The phase diagram for the model in Eq.~(\ref{action}) with $\theta=2\pi/3$. Phase (0) contains no loops. Phase (I) contains proliferated loops in $J_1$ and no loops in $J_2$, while in phase (II) the variables are interchanged. In phase (IV) both species of loops are condensed in single-strength loops. Phase (III) is a condensate of bound states comprised of three charges from one species of loop and a vortex from the other species. The precise description of these condensates is given in the text. } 
\label{phase}
\end{figure}

Figure~\ref{phase} shows the phase diagram for the model with $\theta=2\pi/3$. When both $t_1$ and $t_2$ are small we have a phase [labeled (0) in the figure] where there are only small loops. When $t_1$ is large and $t_2$ is small, we get a phase [labeled (I) in the figure] where one species of loop has proliferated, while the other species has only small loops. Since our model is symmetric under interchange of $t_1$ and $t_2$, we get similar behavior in phase (II). Since these phases do not have both species of loops occurring at the same time, the statistical interactions are unimportant.

We now consider the region of the phase diagram where $t_1$ and $t_2$ are similar, in particular in this work we will often study the ``self-dual'' line where $t_1=t_2$. In this region, if we were to neglect the statistical interaction ($\theta=0$), we would have two phases: a ``gapped'' phase at low $t_1$, $t_2$, where there are only small loops, and a ``condensed'' phase at high $t_1$, $t_2$, with proliferated loops in both the $\vec{J}_1$ and $\vec{J}_2$ variables. The condensed phase would have two gapless modes, one from each species of loop. The transition from the gapped to condensed phase would be two decoupled XY transitions. If we turn on the statistical interaction, we find qualitatively different behavior. For small $t_1$, $t_2$ we again get a gapped phase, but for larger $t_1$, $t_2$, we get a phase, labeled phase (IV) in Fig.~\ref{phase}, where the statistical interactions are manifest more dramatically. We will see below that in this phase both species of loop are condensed, however there are no gapless modes. This phase is distinguished from phase (0) by a non-vanishing correlation between currents of different species. Such correlations were identically zero in the $\theta=\pi$ case, and this phase was not present in that model. 

If we increase $t_1$ and $t_2$ still further, we get a phase, labeled phase (III) in Fig.~\ref{phase}, which is a condensate of bound states composed of three particles in the $\vec{J}_1$ variables and an anti-vortex in the $\vec{J}_2$ variables. This is a (2+1)-dimensional analogue to the $\theta$-term induced ``dyon condensates'' in (3+1) dimensions described in Refs.~\onlinecite{CardyRabinovici1982, Cardy1982, Shapere1989}. Loosely speaking, these composite states appear so that the system can avoid destructive interferences from the statistical interaction.  For example, the statistical interaction in Eq.~(\ref{action}) is inoperative when the $\vec{J}$-currents are present only in multiples of three, while the precise description of the phase (III) is obtained by employing duality approaches in the main text. The transitions from phase (IV) to phases (0) and (III) occur at interesting multicritical points, and we study the system's behavior at these points. 

The outline of this paper is as follows. In Section \ref{sec:model}, we reformulate the model, Eq.~(\ref{action}), in a sign-free way so that we can study it in Monte Carlo. Section \ref{sec:results} contains the results of the Monte Carlo study. These results are presented in terms of the correlation functions of the original $\vec{J}$ variables of Eq.~(\ref{action}), which already allows us to distinguish all phases. In Section \ref{sec:reformulations} we introduce several additional exact reformulations of the model using duality transform\cite{PolyakovBook, Peskin1978, Dasgupta1981, FisherLee1989, LeeFisher1989, artphoton, Fradkin_SL2Z, Witten2003, Rey1991, Lutken1993, Burgess2001} summarized in the Appendix. These reformulations enable us to precisely describe each phase in terms of variables which are gapped in that phase.\cite{LeeFisher1989} This leads us to propose continuum field theories for the various phase transitions in our model in Section \ref{sec:cft}. In Section~\ref{sec:irred} we derive equations for ``irreducible responses'' which provide a physical way to characterize the ``condensates'' that give phases (IV) and (III).  We conclude in Section~\ref{sec:concl} by comparing with the $\theta=\pi$ case and discussing further generalizations.

\section{Monte Carlo Method and Measurements} 
\label{sec:model}
In Ref.~\onlinecite{Geraedts}, we described a method of reformulating models with short-range interactions and statistical terms, such as Eq.~(\ref{action}), in a sign-free way so that they can be studied in Monte Carlo. We review that method here, since in this work we have defined new measurements based on the sign-free reformulation. First, we pass from $J_1$ variables to conjugate 2$\pi$-periodic phase variables by formally writing the constraint at each $r$: 
\begin{equation}
 \delta[\vec{\nabla} \cdot \vec{J_1}(r)=0]=\int_{-\pi}^{\pi} d\phi_{r} \exp[-i \phi_{r} (\vec{\nabla} \cdot \vec{J_{1}})]. 
\label{constraint1}
\end{equation} 
To be precise in our system with periodic boundary conditions, we also require total currents of $\vec{J}_1$ and $\vec{J}_2$ to vanish. In this case we can write $\vec{J}_2=\vec{\nabla} \times \vec{p}_2$ and the action~(\ref{action}) is independent of the gauge choice for $\vec{p}_2$. We enforce the zero total current in $\vec{J}_1$ with the help of fluctuating boundary conditions for the $\phi$-s across a single cut for each direction $\mu=x,y,z$
\begin{equation}
\delta\left[\sum_{r}J_{1\mu}(r)\delta_{r_{\mu},0} \right]=\int_{-\pi}^{\pi} d\gamma_{\mu} \exp\left[-i \gamma_{\mu} \sum_r J_{1\mu}(r) \delta_{r_{\mu},0}\right] .
\label{constraint2}
\end{equation} 

This gives the following partition function:
\begin{equation}
Z= \sum_{{\rm constrained~}\vec{J}_2} \int_{-\pi}^{\pi} \prod_r d\phi_{r} \int_{-\pi}^{\pi} \prod_{\mu=1}^{3}  d\gamma_{\mu} 
e^{-S[\phi,\gamma,\vec{p}_2]}
\label{Z2}
\end{equation}
where the action is given by:
\begin{eqnarray}
S[\phi,\gamma,\vec{p}_2]&&=\sum_r \frac{[\vec{\nabla} \times \vec{p}_2(r)]^2}{2t_2}
\label{action2}\\
&&+\sum_{r,\mu} V_{{\rm Villain}}[\phi_{r+\mu}-\phi_r-\theta p_{2\mu}(r)- \gamma_{\mu} \delta_{r_{\mu},0}].
\nonumber
\end{eqnarray}
$V_{{\rm Villain}}$ is the ``Villain potential'', which is obtained by summing over the $J_1$ variables:
\begin{equation}
\exp[-V_{{\rm Villain}}(\alpha,t_1)]=\sum_{J_{1}=-\infty}^{\infty} \exp \left[-\frac{J_{1}^2}{2t_1}+iJ_1\alpha \right]
\label{Villain}
\end{equation}

In the actual Monte Carlo, we use $\phi_r$, $\gamma_{\mu} \epsilon (-\pi,\pi)$, $p_{2\mu}(r) \epsilon \mathbb{Z}$, and perform unrestricted Metropolis updates. One can show that physical properties measured in such a simulation are precisely as in the above finitely defined model. 

In this work, we monitor ``internal energy per site'', $\epsilon=S/{\rm Vol}$, where ${\rm Vol}=L^3$ is the volume of the system, and compute heat capacity, defined as 
\begin{equation}
C=(\langle \epsilon^2 \rangle - \langle \epsilon \rangle ^2) \times {\rm Vol}.
\label{C}
\end{equation}

To determine the phase diagram, we monitor loop behavior by studying current-current correlations, which are defined as:
\begin{equation}
C^{ab}_{\mu\nu}(k)\equiv\frac{1}{{\rm Vol}}\left \langle J_{a\mu}(k)J_{b\nu}(-k)\right \rangle,
\label{Cgen}
\end{equation}
where $a$ and $b$ are the loop species and $\mu$ and $\nu$ are directions; $J_{a\mu}(k) \equiv \sum_{r} J_{a\mu}(r) e^{-i\vec{k} \cdot \vec{r}}$. 
We trivially have $C^{ba}_{\nu\mu}(k) = C^{ab}_{\mu\nu}(-k)$.
Because of the vanishing total current, we define the correlators at the smallest non-zero $k$; e.g., for $C^{aa}_{xx}$ we used $\vec{k}=(0,\frac{2\pi}{L},0)$ and $\vec{k}=(0,0,\frac{2\pi}{L})$. 
In this work we are interested in the correlations between currents of the same species, $C^{aa}_{\mu\mu}(k)$, also known as the ``superfluid stiffness''.  For example, $C^{22}$ can be measured easily in our Monte Carlo, since we have direct access to $\vec{J}_2 = \vec{\nabla} \times \vec{p}_2$.

We are also interested in the correlations between currents of different species, and we first need to find the corresponding expressions in terms of the variables in Eq.~(\ref{action2}). We can couple the original $\vec{J}$ variables to external probe fields $\vec{A}^{\rm ext}$ by adding the following terms to our action:
\begin{equation}
\delta S= i\sum_{r}\vec{J}_1(r)\cdot\vec{A}^{\rm ext}_1(r)
+ i\sum_{R}\vec{J}_2(R)\cdot\vec{A}^{\rm ext}_2(R).
\label{Aextr}
\end{equation}
We carry the fields $\vec{A}^{\rm ext}_{1,2}$ through the reformulation procedure and then take derivatives of the partition function with respect to them. We obtain the following expression for the correlation between currents of different species:
\begin{eqnarray}
&&C^{12}_{\mu\nu}(k)=\frac{1}{{\rm Vol}} \bigg \langle \left(\sum_{R} J_{2\nu}(R)e^{i\vec{k} \cdot \vec{R}} \right) \nonumber\\
&& \left(i\sum_{r} \frac{\delta V_{{\rm Villain}} (\alpha)}{\delta \alpha} \bigg|_{\nabla_{\mu}\phi-\theta p_{2\mu}-\gamma_{\mu} \delta_{r_{\mu},0}} e^{-i\vec{k} \cdot \vec{r}} \right)  \bigg \rangle.
\label{C12}
\end{eqnarray}
In the above equation, it is important to note that $\vec{J}_1$ and $\vec{J}_2$ are defined on different lattices. In order to work with them on the same footing in $k$-space, it is convenient to define $\vec{R}=\vec{r}^{~\prime}+\vec{d}$, where $\vec{r}^{~\prime}$ is on the direct lattice and $\vec{d}$ is the offset between the two lattices. We can choose any convention for this offset, and we choose $\vec{d}=(1/2,1/2,1/2)$, which means that the sites of the dual lattice are located at the centers of the cubes forming the direct lattice, and we use such ``physical'' coordinates for all sites when defining the Fourier transforms. For a given variable $\vec{W}(r)$ on the lattice whose sites are labeled by indices $r$, the quantity $\vec{\nabla}\times\vec{W}(r)$ is defined on a dual lattice. Now that we have defined the relation between the two lattices, we can precisely define the meaning of the curl operation in $k$-space,
\begin{equation}
[\vec{\nabla}\times\vec{W}]_{\rho}(k)=2i\epsilon_{\rho\nu\mu}\sin(k_{\nu}/2)e^{ik_{\rho}/2}e^{-ik_{\mu}/2}W_{\mu}(k).
\label{curl}
\end{equation}

We can use symmetry arguments to determine some properties of the correlators $C^{ab}_{\mu\nu}(k)$. For simplicity, in this work we define $k$ to be in the $z$-direction, so that $\vec{k}=(0,0,k_z)$, and we only need to consider $\mu,\nu \in \{x,y\}$.  For a symmetry operation to leave our action in Eq.~(\ref{action}) unchanged, it must preserve the relative orientation of two cross-linked currents, like those in Fig.~\ref{loops}. One symmetry that satisfies this requirement involves mirror reflection about a plane while also reversing the direction of one species of loop. $C^{aa}_{xy}(k)$ and $C^{12}_{\mu\mu}(k)$ change sign under such an operation about a plane perpendicular to the $x$-axis, and therefore must be zero.
We can also use such an operation about a plane perpendicular to the $z$-axis to show that $C^{12}_{xy}(k)$ is an odd function of $k$, and hence $C^{21}_{yx}(k) = C^{12}_{xy}(-k) = -C^{12}_{xy}(k)$. Our action is also unchanged if we take its complex conjugate while also reversing the direction of one species of loop. We can use this, along with our precise definition of the offset between the two lattices and of the Fourier transforms, to show that all the correlators $C^{ab}_{\mu\nu}(k)$ are real. Lastly, we can use the $\pi/2$ lattice rotation symmetry of the action to show that $C^{aa}_{xx}(k) = C^{aa}_{yy}(k)$ and $C^{12}_{xy}(k) = -C^{12}_{yx}(k)$. Whenever we present numerical data, we have performed appropriate averages over all directions to improve statistics. 

\section{Results} 
\label{sec:results}
\subsection{Mapping out the phase diagram}
\begin{figure}[t]
\includegraphics[angle=-90,width=\linewidth]{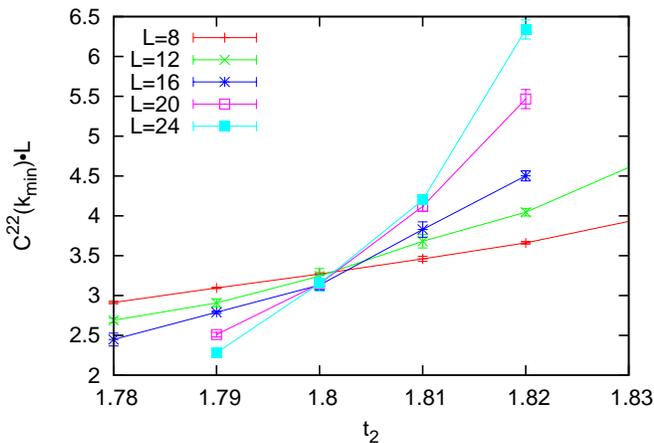}
\caption{A sample of the $C^{22}(k_{\rm min}) \cdot L$ crossings from which the location of the phase transition was determined. Error bars come from comparing runs with different random seeds. Here the transition is from phase (IV) to phase (II). The data is taken for $t_1=1.0$, and we determine the phase transition to be at $t_2=1.80 \pm 0.01$. There is no sign of drift in the location of the crossings, suggesting a second-order transition. Note that the $\vec{J}_2$ variables are condensed in both phases (IV) and (II).  Nevertheless, $C^{22}(k_{\rm min})\cdot L\sim 1/L$ in phase (IV) and $C^{22}(k_{\rm min})\cdot L\sim L$ in phase (II), so the crossing analysis on the plot detects the transition. This subtlety is discussed in the text.}
\label{rhocross}
\end{figure}

We determined the different phases of the model by looking at the stiffness $C^{22}(k)=C^{22}_{xx}(k)=C^{22}_{yy}(k)$, defined at $k=k_{\rm min}\equiv(0,0,2\pi/L)$. Its $L\rightarrow \infty$ limit is non-zero in phases (II) and (III) and vanishes in the other phases. Since our model is exactly symmetric around the self-dual line, we know that $C^{11}(k_{\rm min})$ is non-zero in phases (I) and (III). 
 We found the locations of the phase transitions more accurately by studying $C^{22}(k_{\rm min}) \cdot L$ crossings. We took data in sweeps across the phase diagram (see Fig.~\ref{phase}), and defined the intersection of the $C^{22}(k_{\rm min}) \cdot L$ curves to be the location of the phase transition. An example of such a sweep is shown in Fig.~\ref{rhocross}. The dots on the phase diagram in Fig.~\ref{phase} are the locations of the phase transitions determined in this way. In all $C^{22}(k_{\rm min}) \cdot L$ sweeps, we found that the crossings did not drift with increasing $L$, which suggests that these phase transitions are second-order.

Let us now consider some limiting cases. The model with $t_1=0$ is a model containing only one species of loop.\cite{Cha1991} Our value for the position of the (0)-(II) XY transition ($t_2 \approx 0.333...$) is in agreement with prior work on this model.\cite{Sorensen} The transition is in the 3D XY universality class also for small, non-zero $t_1$. 

For $t_1 \rightarrow \infty$, the Villain weight (\ref{Villain}) vanishes except for $\alpha=2\pi \times$ (int), which enforces $\vec{J}_2=\vec{\nabla} \times \vec{a}_2=3\times$(int). Therefore, at $t_1 \rightarrow \infty$ the (I)-(III) transition is a transition from no loops of $J_2$ to $J_2$ loops of strength $3$. One expects that this transition is XY-like, and similar to the (0)-(II) transition, but due to tripled $J_2$, it should occur at a $t_2$ value nine times higher. We observed the (I)-(III) transition to occur at $t_2 \approx 3$ for large $t_1$, in agreement with this expectation. We give a precise description of phase (III) for finite $t_1$, $t_2$ in Sec. \ref{sec:reformulations}. 

\begin{figure}[t]
\includegraphics[angle=-90,width=\linewidth]{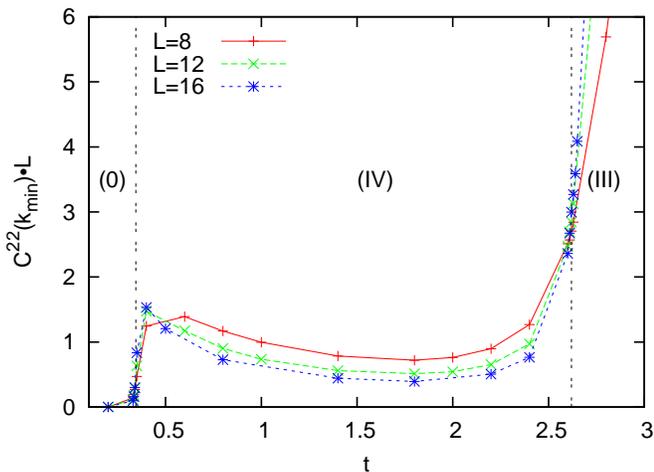}
\caption{$C^{22}(k_{\rm min})\cdot L$ along the self-dual line. We can see that $C^{22}(k_{\rm min})\cdot L$ vanishes in phases (0) and (IV), despite its unusual behavior at the transition between these phases. $C^{22}(k_{\rm min})\cdot L$ diverges in phase (III) due to the proliferation of triple-strength loops in the $\vec{J}$ variables. The vertical lines mark the locations of the phase transitions. More detailed data at the phase transitions is shown in Figs.~ \ref{R41} and \ref{R42}.  }
\label{C22}
\end{figure}

\begin{figure}[t]
\includegraphics[angle=-90,width=\linewidth]{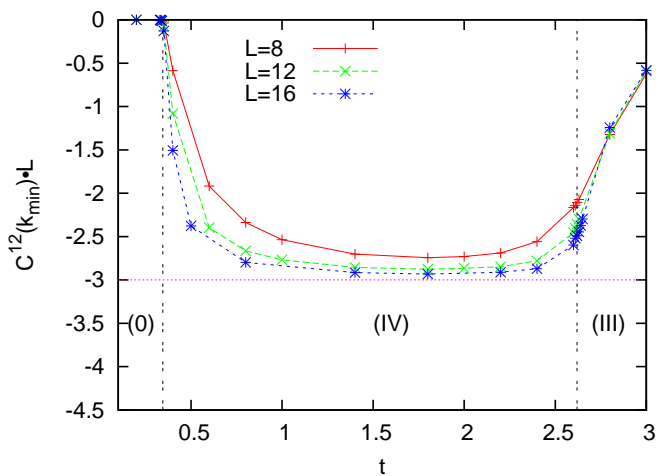}
\caption{$C^{12}(k_{\rm min}) \cdot L$ along the self-dual line. $C^{12}(k_{\rm min}) \cdot L$ vanishes in phase (0), but it approaches a universal value in phase (IV). We predict this value to be at $-2\pi/\theta=-3$, which is shown by a horizontal line. In phase (IV), $C^{12}(k_{\rm min}) \cdot L$ approaches a non-universal value. The vertical lines mark the locations of the phase transitions.}
\label{fig:C12}
\end{figure}

In Fig.~\ref{C22} we show $C^{22}(k_{\rm min})\cdot L$ along the self-dual line $t_1=t_2$, going through phases (0), (IV), and (III). Neither of phases (0) and (IV) have a finite superfluid stiffness $C^{22}(k_{\rm min})$, so to distinguish between them we use the correlator $C^{12}_{xy}(k_{\rm min})$, denoted as $C^{12}(k_{\rm min})$ in what follows. A plot of $C^{12}(k_{\rm min})$ along the self-dual line is shown in Fig.~\ref{fig:C12}. $C^{12}(k_{\rm min})\cdot L$ vanishes in phase (0) in the $L\rightarrow \infty$ limit, but is non-zero in phase (IV), so the two phases are indeed different.

We can understand the observations in phases (0) and (IV) as follows. The excitations in phase (0) are small loops in the $\vec{J}$ variables, which implies that in this phase $C^{aa}(k) \sim k^2$ for small $k$. For $k=k_{\rm min}$, this gives $C^{22}(k_{\rm min}) \sim 1/L^2$. The smallest excitation that contributes to $C^{12}$ consists of a small loop in each of the $\vec{J}_1$ and $\vec{J}_2$ variables. An estimate of such contributions with cross-linking between the loops leads to $C^{12}(k) \sim -\sin(\theta)k^3$.

In phase (IV), the $\vec{J}$ variables are condensed. One way of expressing this condensation is to replace the integer-valued $\vec{J}$ with real-valued variables $\vec{j}$. This is equivalent to coarse-graining the model and integrating out the gapped vortices (see Sec.~\ref{sec:cft}). If we define new gauge variables $\vec{\alpha}_{j1}$ and $\vec{\alpha}_{j2}$ such that $\vec{j}_1 = \frac{\vec{\nabla} \times \vec{\alpha}_{j1}}{2\pi}$ and $\vec{j}_2 = \frac{\vec{\nabla} \times \vec{\alpha}_{j2}}{2\pi}$, then we can replace the original action Eq.~(\ref{action}) by an effective action in terms of the $\vec{\alpha}_{j1}$, $\vec{\alpha}_{j2}$ variables:
\begin{eqnarray}
S_{\rm eff}[\vec{\alpha}_{j1},\vec{\alpha}_{j2}]&=&\frac{1}{2} \sum_R \frac{[\vec{\nabla} \times \vec{\alpha}_{j1}(R)]^2}{(2\pi)^2t_{1, {\rm eff}}}+\sum_r\frac{[\vec{\nabla} \times \vec{\alpha}_{j2}(r)]^2}{(2\pi)^2t_{2,{\rm eff}}}\nonumber\\
&+&\frac{i\theta}{(2\pi)^2}\sum_{r} [\vec{\nabla} \times \vec{\alpha}_{j1}(R)] \cdot \vec{\alpha}_{j2}(r).
\label{CSJ}
\end{eqnarray}
In the absence of the last ``mutual Chern-Simons'' (CS) term, this would be an action for two decoupled gauge fields, which would have two gapless modes. When the mutual CS term is included, it destroys the gapless modes. We can calculate the $C^{aa}(k)$ and $C^{12}(k)$ correlators with respect to this gaussian action, and we find that $C^{aa}(k) \sim k^2\sim1/L^2$ for $k=k_{\rm min}$, consistent with our data. We also find that $C^{12}(k)\approx-\frac{k}{\theta}=-\frac{2\pi}{L\theta}=-\frac{3}{L}$ for $k=k_{\rm min}$. This quantity is represented by the dotted line in Fig.~\ref{fig:C12}, and we can see that our Monte Carlo data approach this value.

Let us briefly remark on the use of $C^{22}(k_{\rm min})\cdot L$ crossings to determine the phase boundaries. It is natural to use these crossings on the (0)-(II) and (I)-(III) transitions, where we are going from a phase with only small $\vec{J}_2$ loops to a phase with large $\vec{J}_2$ currents. For the transition from phase (IV) to (II), we are going between two phases where the $\vec{J}_2$ variables are condensed. However, since in phase (IV) $C^{22}(k_{\rm min})\cdot L \sim 1/L$ while in phase (II) $C^{22}(k_{\rm min})\cdot L \sim L$ we can still use crossings in this quantity to determine the transition between the two phases. One might not expect, however, to see qualitatively similar behavior between this transition and the transitions (0)-(II) and (I)-(III), yet this is what we observe. The reasons for this will be explained in Sec.~\ref{sec:reformulations}. For the transition between phase (I) and phase (IV), $C^{22}(k_{\rm min}) \cdot L\sim 1/L$ in both phases, and so we cannot use it to detect this transition. 

\subsection{Transition (0)-(IV) along the self-dual line}
\begin{figure}[t]
\includegraphics[angle=-90,width=\linewidth]{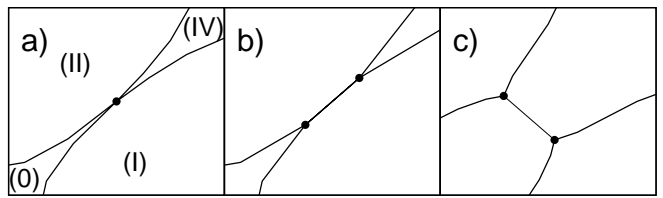}
\caption{Different scenarios for how the phases can meet on the self-dual line. In scenario a), all four phases meet at a point. In b), phases (I) and (II) meet on a line segment on the self-dual line, and in c) phases (0) and (IV) meet on a line segment perpendicular to the self-dual line. We believe that scenario c) is unlikely, but we cannot distinguish between a) and b) at finite size. We can only state that if such a segment exists in scenario b), it is in the narrow range $t=0.335-0.35$.}
\label{scenarios}
\end{figure}

\begin{figure}[t]
\includegraphics[angle=-90,width=\linewidth]{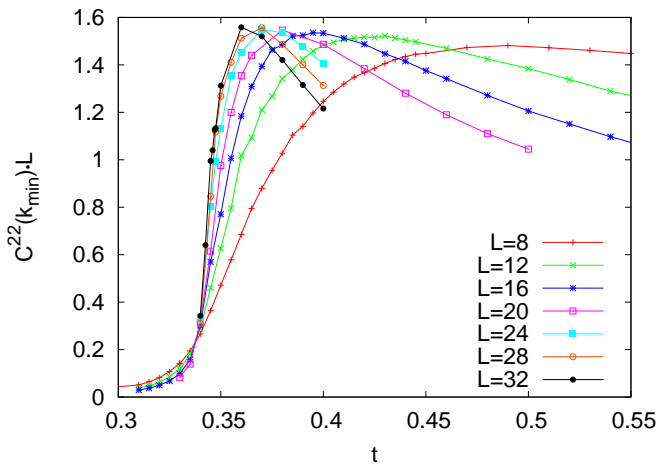}
\caption{$C^{22}(k_{\rm min}) \cdot L$ along the self-dual line at the bottom multicritical point. $C^{22}(k_{\rm min}) \cdot L$ vanishes as $1/L$ for sufficiently large $L$ in both phase (0) and phase (IV), but is non-zero at the transition.}
\label{R41}
\end{figure}

We now investigate the apparent multicritical points on the self-dual line. We first study the lower regime where phases (0), (IV), (I) and (II) meet.  
We are interested in how the phases join. Due to the symmetry between $t_1$ and $t_2$, there are three scenarios, shown in Fig.~\ref{scenarios}. In a), all four phases meet at a point, while in b) there is a critical line segment on the self-dual line, and in c) such a segment is perpendicular to the self-dual line. Figure~\ref{R41} shows $C^{22}(k_{\rm min}) \cdot L$ along the self-dual line near this transition. $C^{22}(k_{\rm min}) \cdot L$ vanishes in phases (0) and (IV) [see also Fig.~\ref{C22}], but appears to have a finite value in the critical regime. If scenario c) were correct, we would expect $C^{22}(k_{\rm min}) \cdot L$ to vanish at the (0)-(IV) transition since phases (I) and (II) should not influence its behavior. In addition, we have taken sweeps with $t_2=t_1+\delta t, \delta t=0.002$, which are lines parallel to the self-dual line and displaced from it by $\delta t$. We found two distinct phase transitions near the critical point, so if scenario c) is accurate the line segment is $<0.004$ in size. For these reasons, we believe that scenario c) is not taking place. 

\begin{figure}[t]
\includegraphics[angle=-90,width=\linewidth]{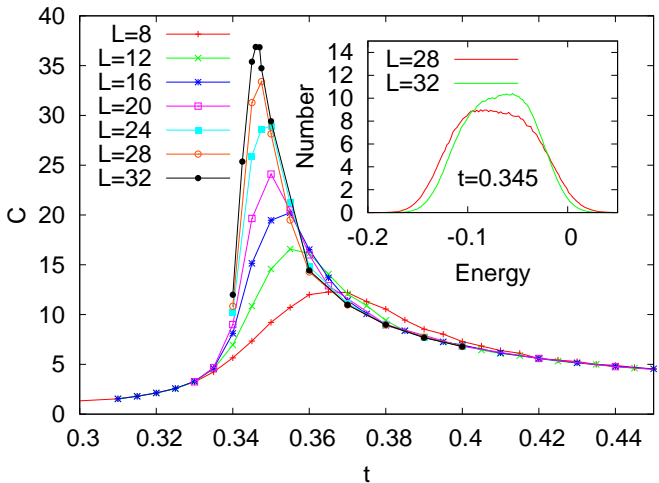}
\caption{Heat capacity along the self dual line at the bottom multicritical point. The sharpness of the peaks suggests either a small $\nu$ or a first-order phase transition between phases (0) and (IV). The inset shows histograms of the energy per site $\epsilon$, at $L=28,32$, $t=0.345$. The irregular shape of the histogram suggests that this is a first-order transition.}
\label{C41}
\end{figure}

Furthermore, if there is a line segment as in scenario b), it is no larger than the region in Fig.~\ref{R41} where $C^{22} \cdot L$ is increasing with system size. We can further limit this segment by studying heat capacity shown in Fig.~\ref{C41}, and noting that the segment is no larger than the region where heat capacity increases with system size. We therefore estimate that the line segment is within the small range $t\in [0.335$, $0.35$]. Studying larger sizes could further narrow our bounds on the possible extent of the line segment, but at finite system size we cannot show that it does not exist. 

To determine the order of the transition at this point, we studied how the heat capacity increases with system size. We can see from Fig.~\ref{C41} that the heat capacity has a sharp peak around $t=0.345$. We also studied histograms of the total energy per site $\epsilon$. In the second-order case, these histograms would be singly-peaked, while in the first-order case we expect to see two distinct peaks. An example of such a histogram analysis, taken at $t=0.345$, $L=28$ and $32$, is shown in the inset for Fig.~\ref{C41}. We do not see two distinct peaks, however the histograms have a ``flat top'', suggestive of two peaks which are too close to be distinguishable on our finite sizes. This flat top suggests that we have a first-order transition. 

The data we have presented seems to suggest that we have a first-order transition in the form of scenario a), which would be highly unusual. We therefore propose two alternate scenarios. Firstly, the transition could be first-order and scenario b), with a line segment which is very small. However, in addition to the small size of the line segment, if this scenario were true we would expect $C^{22}(k_{\rm min})$ to be finite at the transition, since it is non-zero in phase (II). Therefore we would expect $C^{22}(k_{\rm min})\cdot L$ to increase linearly with $L$ on the segment, and this is not consistent with Fig.~\ref{R41}. Alternatively, scenario a) could indeed be correct, and the transition could be second order. The behavior of $C^{22}(k_{\rm min})\cdot L$ in Figs.~\ref{C22} and \ref{R41} implies that we have very strong crossovers in our simulation variables: as we approach the transition from phase (IV), we need larger and larger sizes to see the eventual vanishing of $C^{22}(k_{\rm min})\cdot L\sim 1/L$ in this phase. It is possible that the unusual shape of the energy histograms could be due to sampling in these variables. Studying the system at larger sizes could help to resolve these questions, by both more clearly resolving the histograms and further reducing the extent of the possible line segment of scenario b).
 
\subsection{Transition (IV)-(III) along the self-dual line}
\begin{figure}[t]
\includegraphics[angle=-90,width=\linewidth]{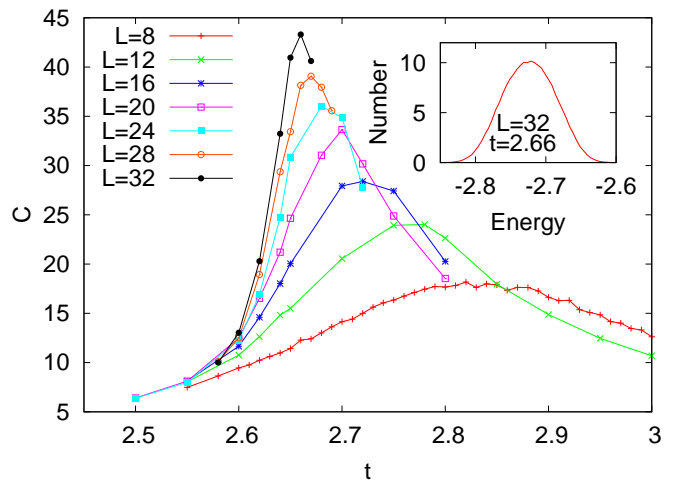}
\caption{Heat capacity along the self-dual line at the upper multicritical point. The behavior of the peak with size suggests a second-order transition. The inset shows a histogram of the energy per site $\epsilon$, at $L=32, t=2.66$, which is where $C$ has a maximum for this size. The single peak further suggests a second-order transition. We also studied histograms at all of the other points where heat-capacity was measured, and found only single-mode distributions everywhere.}
\label{C42}
\end{figure}

\begin{figure}[t]
\includegraphics[angle=-90,width=\linewidth]{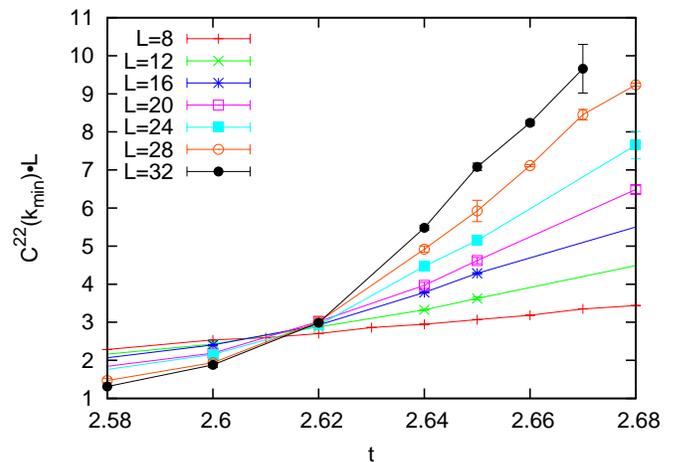}
\caption{$C^{22}(k_{\rm min}) \cdot L$ along the self-dual line at the upper multicritical point. The crossings do not drift with increasing size, implying a second-order transition at $t_{\rm crit}=2.62$.}
\label{R42}
\end{figure}

Figure~\ref{C42} shows the heat capacity in the regime where phases (IV), (III), (I), and (II) meet. The peaks in the heat capacity evolve only slowly with system size, suggesting a second-order phase transition. Figure~\ref{R42} shows the $C^{22}(k_{\rm min}) \cdot L$ near this point. At this transition we are going from a phase with $C^{22}(k_{\rm min}) \cdot L \sim 1/L$ to $C^{22}(k_{\rm min}) \cdot L \sim L$, so we expect a crossing at the phase transition. We observe that this crossing does not drift with increasing $L$, further supporting the conclusion that the transition is second order. Finite-size scaling arguments suggest that $C^{22}(k_{\rm min}) \cdot L =f[(t-t_{\rm crit})L^{1/\nu}]$ in our model. We can therefore try to collapse the $C^{22}(k_{\rm min}) \cdot L$ data in Fig.~\ref{R42} to one curve by  rescaling the horizontal axis by $(t-t_{\rm crit})L^{1/\nu}$. Applying this process, using $t_{\rm crit}=2.62$ inferred from Fig.~\ref{R42}, gives $\nu=0.8 \pm 0.1$, consistent with a second-order transition. We have also obtained histograms of total energy at all of the points in Fig.~\ref{C42}, and have found singly-peaked histograms at all points. The inset to Fig.~\ref{C42} shows our data for $t=2.66, L=32$, which is the location of the heat capacity maximum in the figure. This phase transition is a transition from a phase where the $\vec{J}$ variables are condensed in single strengths to a phase where they are proliferated only in triple strengths. However, we have used techniques for analyzing $C^{22}(k_{\rm min}) \cdot L$ which are valid for the ordinary condensation of loop variables. This will be justified by a more precise description of the two phases and the transition between them in Secs.~\ref{sec:reformulations} and~\ref{sec:cft}. 

\section{Analysis in terms of exact reformulations}
\label{sec:reformulations}
Using the duality transform shown in the Appendix, we can derive several exact reformulations of the action in Eq.~(\ref{action}). We can use these reformulations to describe each phase in terms of variables whose loops are gapped in that phase. The nature of the different phases and the transitions between them can then be described in terms of these variables. We can also introduce into our initial action external ``probe'' gauge fields coupled to $\vec{J}_1$ and $\vec{J}_2$, by adding terms to the action similar to those in Eq.~(\ref{Aextr}):
\begin{equation}
\delta S = i\sum_k \vec{J}_1(-k)\cdot\vec{A}^{\rm ext}_1(k)
+ i\sum_k \vec{J}_2(-k)\cdot\vec{A}^{\rm ext}_2(k).
\label{Aext}
\end{equation}
 We can carry these gauge fields through the duality transforms as illustrated in the Appendix, and take derivatives with respect to them to obtain various exact relations between different current-current correlators. We will use such relations to better understand the behavior of these correlators.

To obtain an action suitable for describing phase (I), we apply the duality procedure outlined in the Appendix to the $\vec{J}_1$ variables in our initial action. We obtain the following reformulation:\cite{footnotetwo}
\begin{widetext}
\begin{eqnarray}
S[\vec{Q}_1, \vec{J}_2] &=& \frac{1}{2} \sum_k \frac{|2\pi\vec{Q}_1(k) + \theta(k) \vec{J}_2(k)|^2}{v_1(k) |\vec{f}_k|^2} + \frac{1}{2} \sum_k v_2(k) |\vec{J}_2(k)|^2 \nonumber\\
&=&\frac{1}{2} \sum_k \left[\frac{(2\pi)^2}{v_1(k)|\vec{f}_k|^2} |\vec{Q}_1(k)|^2 + \left(v_2(k) + \frac{\theta(k)^2}{v_1(k)|\vec{f}_k|^2} \right) |\vec{J}_2(k)|^2 + \frac{4\pi\theta(k)}{v_1(k)|\vec{f}_k|^2} \vec{Q}_1(-k) \cdot \vec{J}_2(k) \right] ~, 
\label{SJQ}
\end{eqnarray}
\end{widetext}
where $f_{k,\mu}\equiv1-e^{ik_{\mu}}$, as defined in the Appendix. The action is written in terms of $\vec{J}_2$ variables, and  $\vec{Q}_1$ variables that are dual ``vortex'' variables to $\vec{J}_1$. In the above action, and from now on, we consider the case of general intra-species interactions $v_1(k)$ and $v_2(k)$, though in the preceding section we considered specific short-range interactions $v_1(k) \equiv 1/t_1$ and $v_2(k)\equiv 1/t_2$. We also consider a more general, $k$-dependent statistical coefficient $\theta(k)$. 
Throughout this work, we will assume that $v_1(k),v_2(k)$ and $\theta(k)$ are real and satisfy $v_a(k)=v_a(-k)$ and $\theta(k)=\theta(-k)$. 
We can see in the above action that for large $t_1$ and small $t_2$, both $\vec{Q}_1$ and $\vec{J}_2$ have a large energy cost, and therefore both are gapped. We expect this; since the $\vec{J}_1$ variables are condensed, the variables dual to them should be gapped. Naturally, we can obtain a reformulation for phase (II) by applying the same steps to the $\vec{J}_2$ variables.

To get an action suitable for describing phase (IV), we apply the duality procedure to the $\vec{J}_2$ variables in Eq.~(\ref{SJQ}). This gives us the following action, expressed in terms of ``vortex'' variables $\vec{Q}_1$ and $\vec{Q}_2$ that are dual to the $\vec{J}_1$ and $\vec{J}_2$ variables:
\begin{widetext}
\begin{eqnarray}
S[\vec{Q}_1, \vec{Q}_2] = \frac{1}{2} \sum_k \frac{(2\pi)^2 \left[v_2(k) |\vec{Q}_1(k)|^2 + v_1(k) |\vec{Q}_2(k)|^2 \right]}{|\vec{f}_k|^2 v_1(k) v_2(k) + \theta(k)^2} - i\sum_k \frac{(2\pi)^2 \theta(k) ~~ \vec{Q}_1(-k) \cdot \vec{p}_{Q2}(k)}{|\vec{f}_k|^2 v_1(k) v_2(k) + \theta(k)^2} ~.
\label{Sdd}
\end{eqnarray}
\end{widetext}
Here $\vec{p}_{Q2}$ is an auxiliary ``gauge field'' encoding the flux of $\vec{Q}_2$, and is defined such that $\vec{Q}_2 = \vec{\nabla} \times \vec{p}_{Q2}$ (because of the constraints on $\vec{Q}_{1,2}$, the action is independent of the choice of $\vec{p}_{Q2}$). Unlike the analysis of phase (I) in the $\vec{Q}_1$, $\vec{J}_2$ variables, it is not clear that a phase with gapped $\vec{Q}_1$, $\vec{Q}_2$ exists. If we define $v_{1/2,{\rm dual}}$ such that 
\begin{equation}
v_{1/2,{\rm dual}}(k)=\frac{(2\pi)^2v_{2/1}(k)}{|\vec{f}_k|^2v_1(k)v_2(k)+\theta(k)^2},
\label{vdual}
\end{equation}
we see that $v_{1/2,{\rm dual}}$ cannot both be arbitrarily large, and so the interactions may not be large enough to gap out both $\vec{Q}_1$ and $\vec{Q}_2$. Considering comparable $v_1 \sim v_2$, the dual interactions are largest for intermediate $v_1$ and $v_2$ and their magnitude increases with decreasing $\theta$. Whether a phase with both species gapped exists needs to be determined numerically. We have found that in the current model with $\theta=2\pi/3$, phase (IV) is the phase where $\vec{Q}_1$ and $\vec{Q}_2$ are gapped. In contrast, in the $\theta=\pi$ model,\cite{Geraedts} such a phase did not exist, and instead we expect either a critical state or phase separation.\cite{Senthil2006_theta,Xu2011}

We also note that the transition from phase (IV) to phase (II) is a transition where the $\vec{Q}_1$ variables are going from gapped to condensed, while the $\vec{Q}_2$ variables remain gapped. Therefore, if we could study correlators such as $\langle Q_{1\mu}(k)Q_{1\mu}(-k)\rangle $, we would expect them to behave in the well-understood manner of one species of loop condensing, qualitatively similar to the behavior of the $\vec{J}_2$ variables in the (0)-(II) transition. We now invoke a useful relation derived by introducing external probe fields as explained above:
\begin{widetext}
\begin{eqnarray}
\label{QtoJ2}
&&C^{22}_{yy}[(k=(0,0,k_z)] =\frac{v_1(k)|\vec{f}_k|^2}{|\vec{f}_k|^2v_1(k)v_2(k)+\theta(k)^2}\\
&&+\frac{(2\pi)^2\left[\theta(k)^2\langle Q_{1y}(k)Q_{1y}(-k)\rangle -|\vec{f}_k|^2v_1(k)^2\langle Q_{2x}(k)Q_{2x}(-k)\rangle +4\sin(k_z/2)v_1(k)\theta(k)\langle Q_{1y}(k)Q_{2x}(-k)\rangle\right] }{\left[|\vec{f}_k|^2v_1(k)v_2(k)+\theta(k)^2\right]^2}.\nonumber
\end{eqnarray}
\end{widetext}
$\vec{Q}_2$ is gapped everywhere near the (IV)-(II) phase transition, which implies that the excitations of $\vec{Q}_2$ are small loops, and we can show that $\langle Q_{2\mu}(k)Q_{2\mu}(-k) \rangle \sim k^2$ for small $k$ in the region of the transition. We also expect that $\langle Q_{1y}(k)Q_{2x}(-k) \rangle \sim k^3$ in phase (IV) and $\sim k$ in phase (II). Taking the limit of small $k$, we see that most of the terms are of order $k^2$ or smaller, and we are left with
\begin{equation}
C^{22}_{yy}(k)=\frac{(2\pi)^2}{\theta(k)^2}\langle Q_{1y}(k)Q_{1y}(-k)\rangle +O(k^2).
\label{Q2J}
\end{equation}
This explains why $C^{22}(k_{\rm min})\sim 1/L^2$ in phase (IV) and is constant in phase (II), even though $\vec{J}_2$ is condensed in both phases. It allows us to use $C^{22}$ to study the condensation of the $\vec{Q}_1$ variables, which is what we showed in Fig.~\ref{rhocross}. Equation~(\ref{Q2J}) is also valid at the transition between phase (IV) and phase (III), shown in Fig.~\ref{R42}. We can see from this figure that the $\vec{Q}$ variables seem to be undergoing a continuous transition; we will discuss this further in Sec.~\ref{subsec:IV-III}.

We can also establish the behavior of $C^{12}_{xy}(k)$ in phase (IV) by invoking another relation, again derived by differentiating the partition function with respect to external probe fields:
\begin{widetext}
\begin{eqnarray}
&&C^{12}_{xy}(k)=-\frac{2\sin(k_z/2)\theta(k)}{|\vec{f}_k|^2v_1(k)v_2(k)+\theta(k)^2}
+\frac{(2\pi)^22\sin(k_z/2)\theta(k)}{\left[|\vec{f}_k|^2v_1(k)v_2(k)+\theta(k)^2\right]^2}
\left[ v_1(k)\langle Q_{2x}(k)Q_{2x}(-k)\rangle+v_2(k)\langle Q_{1y}(k)Q_{1y}(-k)\rangle\right]\nonumber\\
&&~~~~~~~+\frac{(2\pi)^2\left[|\vec{f}_k|^2v_1(k)v_2(k)-\theta(k)^2\right]}{\left[|\vec{f}_k|^2v_1(k)v_2(k)+\theta(k)^2\right]^2}\langle Q_{1y}(k)Q_{2x}(-k)\rangle.
\label{C12inQ}
\end{eqnarray}
\end{widetext}
We can see that for gapped $\vec{Q}_{1,2}$ and in the small $k$ limit, $C^{12}_{xy}$ indeed approaches $-k/\theta$, as we argued from a schematic treatment of the $\vec{J}_1$, $\vec{J}_2$ condensate in Eq.~(\ref{CSJ}). 
 Note that we can use Eqs.~(\ref{QtoJ2}) and (\ref{C12inQ}) to express the $\vec{Q}$ correlators in terms of the $\vec{J}$ correlators. We have done this, and plots of the data (not shown), confirm the condensation of $\vec{Q}_1$ across the (IV)-(II) and (IV)-(III) transitions, as expected from the above analysis.  We chose to express all data in terms of correlators in the $\vec{J}$ variables so that Section~\ref{sec:results} could be understood without any knowledge of the various reformulations. 

We can also give precise meaning to the treatment in Eq.~(\ref{CSJ}). From the Appendix, an intermediate step in the duality procedure going from $\vec{J}_1,\vec{J}_2$ to $\vec{Q}_1,\vec{Q}_2$ is:
\begin{widetext}
\begin{eqnarray}
S[\vec{\alpha}_{j1},\vec{\alpha}_{j2},\vec{Q}_1,\vec{Q}_2]&=&\frac{1}{2} \sum_{R} \frac{[\vec{\nabla} \times \vec{\alpha}_{j1}(R)]^2}{(2\pi)^2t_1}+\sum_r\frac{[\vec{\nabla} \times \vec{\alpha}_{j2}(r)]^2}{(2\pi)^2t_2}\nonumber+\frac{i\theta}{(2\pi)^2}\sum_{r} [\vec{\nabla} \times \vec{\alpha}_{j1}(R)] \cdot \vec{\alpha}_{j2}(r)\nonumber \\
&&+i\sum_R \vec{Q}_1(R)\cdot\vec{\alpha}_{j1}(R)+i\sum_r \vec{Q}_2(r)\cdot\vec{\alpha}_{j2}(r).
\label{CSJ2}
\end{eqnarray}
\end{widetext}
Gaussian integration over the $\vec{\alpha}$ variables gives Eq.~(\ref{Sdd}). Equation (\ref{CSJ2}) is an action for two gauge fields with mutual Chern-Simons interactions coupled to integer-valued currents $\vec{Q}$. When $\vec{Q}_1$ and $\vec{Q}_2$ are gapped, we can formally integrate them out and obtain the low-energy field theory description in Eq.~(\ref{CSJ}). 

\begin{figure}[t]
\includegraphics[width=0.6\linewidth]{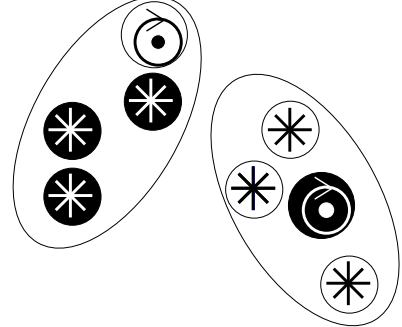}
\caption{Illustration of the ``molecules'' that are condensed in phase (III). Each molecule contains three charges (stars), and one anti-vortex (circles). Black objects on white backgrounds are of species 1, and white objects on black backgrounds are of species 2.}
\label{bound}
\end{figure}

We now consider a reformulation appropriate for the description of phase (III). Our crude intuition is that the $\vec{J}_1$ and $\vec{J}_2$ loops will indeed condense strongly but only in multiples of $n$, where $\theta=\frac{2\pi}{n}$, in order to avoid the statistical interaction. To proceed more accurately, we start with $S[\vec{Q}_1, \vec{J}_2]$, Eq.~(\ref{SJQ}), and notice that for small $v_1$ and $v_2$ the combination 
\begin{equation}
\vec{M}_2(R) \equiv \vec{J}_2(R) + n \vec{Q}_1(R)
\label{M2}
\end{equation}
wants to be gapped while $\vec{Q}_1$ wants to be ``condensed'', hence $\vec{J}_2$ wants to be ``condensed'' in multiples of $n$.  More precisely, in the partition sum, we can change the summation variables from integer-valued currents $\vec{Q}_1$ and $\vec{J}_2$ to integer-valued currents $\vec{Q}_1$ and $\vec{M}_2$, with the action
\begin{eqnarray}
&& S^{(\theta=2\pi/n)}[\vec{Q}_1, \vec{M}_2] = \frac{1}{2} \sum_k \frac{(2\pi)^2 |\vec{M}_2(k)|^2}{n^2 v_1(k)|\vec{f}_k|^2} \\
&& ~~~~~~~~~~~ + \frac{1}{2} \sum_k v_2(k) |\vec{M}_2(k) - n\vec{Q}_1(k)|^2 ~.
\end{eqnarray}
We can now consider a phase with $\vec{M}_2$ gapped and $\vec{Q}_1$ condensed appropriate for small $v_1$ and $v_2$.  The precise meaning of the $\vec{Q}_1$ condensation is again obtained by going from $\vec{Q}_1$ to dual variables $\vec{M}_1$ using the formal prescription in the Appendix.  The result is
\begin{widetext}
\begin{eqnarray}
S^{(\theta=2\pi/n)}[\vec{M}_1, \vec{M}_2] = \sum_k \frac{(2\pi)^2}{n^2 |\vec{f}_k|^2} \left[\frac{|\vec{M_1}(k)|^2}{2 v_2(k)} + \frac{|\vec{M}_2(k)|^2}{2 v_1(k)} \right]
+ i \sum_k \frac{2\pi}{n} \vec{M}_1(-k) \cdot \vec{p}_{M2}(k) ~,
\label{Smm}
\end{eqnarray}
\end{widetext}
where $\vec{M}_2 = \vec{\nabla} \times \vec{p}_{M2}$.
Note that if we were to dualize $\vec{Q}_1$ in $S[\vec{Q}_1, \vec{J}_2]$, we would of course obtain back $S[\vec{J}_1, \vec{J}_2]$ (up to sign of $\vec{J}_1$), while the duality procedure after the change of variables in Eq.~(\ref{M2}) gives a different reformulation since here we dualize $\vec{Q}_1$ while keeping $\vec{M}_2$ as an independent current.
Labels $1$ and $2$ on $\vec{M}_1$ and $\vec{M}_2$ are somewhat arbitrary, as we mixed the original species $1$ and $2$, e.g., when defining $\vec{M}_2$ in Eq.~(\ref{M2}).  We can think about a phase with gapped $\vec{M}_2$ as having binding of $n$ original $\vec{J}_2$ currents to one anti-vortex in $\vec{J}_1$, so that for $J_2=n$, $Q_1=-1$, we have $M_2=0$. In phase (III) it is such ($J_2=n$, $Q_1=-1$) molecules that are condensed. These bound states are illustrated in Fig.~\ref{bound}. This is the more accurate description of phase (III), made precise by the reformulation Eq.~(\ref{Smm}) with gapped $\vec{M}_{1,2}$. For $n=2$, this is also the precise description of phase (III) in Fig. 1 of the $\theta=\pi$ statistics model studied previously.\cite{Geraedts} We can treat a small loop of $\vec{M}_2$ as an excitation out of such a bound state. The symmetric structure of the above action suggests a similar interpretation of $\vec{M}_1$ even though this field was introduced differently.

Since the $\vec{M}$ variables are gapped in phase (III), we know from small loop arguments that $\langle M_{a\mu}(k)M_{a\mu}(-k)\rangle  \sim k^2$ and $\langle M_{1\mu}(k)M_{2\nu}(-k)\rangle  \sim k^3$. We can also derive the following exact relations at $k=(0,0,k_z)$:
\begin{eqnarray}
&&C^{22}_{yy}=\frac{1}{v_2(k)}-\frac{(2\pi)^2}{n^2|\vec{f}_k|^2v_2(k)^2}\langle M_{1x}(k)M_{1x}(-k)\rangle \\
&&C^{12}_{xy}=\frac{(2\pi)^2}{n^2v_1(k)v_2(k)|\vec{f}_k|^2}\langle M_{1x}(k)M_{2y}(-k)\rangle 
\end{eqnarray}
These imply that $C^{22}(k)\sim$(constant) and $C^{12}(k) \sim k $. Note that unlike in phase (IV), $C^{12}(k_{\rm min})\cdot L$ has a non-universal value in phase (III), as can be seen in Fig.~\ref{fig:C12}. 

\section{Field Theories for the Phases and Phase Transitions}
\label{sec:cft}
\subsection{Phase (0) and the (0)$\rightarrow$(IV) transition}
Equation~(\ref{CSJ}) is a continuum field theory useful for describing phase (IV). In this phase, the $\vec{J}$ variables are condensed which allowed us to use real-valued variables $\vec{j}=\frac{\vec{\nabla}\times\vec{\alpha}_j}{2\pi}$. In phase (0), the $\vec{Q}$ variables are condensed, and we can replace them with real-valued variables $\vec{q}$. We can then write a field theory in terms of real-valued gauge fields $\vec{\alpha}_q$ such that $\vec{q}=\frac{\vec{\nabla}\times\vec{\alpha}_q}{2\pi}$. Performing this procedure on Eq.~(\ref{Sdd}) gives:
\begin{widetext}
\begin{eqnarray}
\label{CSQ}
S[\vec{\alpha}_{q1},\vec{\alpha}_{q2},\vec{J}_1,\vec{J}_2]&=&\frac{1}{2}\sum_k \left[\frac{v_2(k)|[\vec{\nabla} \times \vec{\alpha}_{q1}](k)|^2}{|\vec{f}_k|^2v_1(k)v_2(k)+\theta(k)^2} +\frac{v_1(k)|[\vec{\nabla} \times \vec{\alpha}_{q2}](k)|^2}{|\vec{f}_k|^2v_1(k)v_2(k)+\theta(k)^2}\right] \\
&-&i\sum_k \frac{\theta(k)}{|\vec{f}_k|^2v_1(k)v_2(k)+\theta(k)^2} [\vec{\nabla} \times \vec{\alpha}_{q1}](-k) \cdot \vec{\alpha}_{q2}(k) 
+i \sum_k \left[\vec{J}_1(-k)\cdot\vec{\alpha}_{q1}(k) + \vec{J}_2(-k)\cdot\vec{\alpha}_{q2}(k)\right], \nonumber
\end{eqnarray}
\end{widetext}
which can be viewed as an intermediate step in the (exact) duality map from the variables $\vec{Q}_{1,2}$ to $\vec{J}_{1,2}$, cf.~the Appendix. 

 We can now take the long-wavelength limit and write a schematic action in real space:
\begin{widetext}
\begin{eqnarray}
\label{CSQft}
&&S_{\rm eff}[\vec{\alpha}_{q1},\vec{\alpha}_{q2},\Psi_{J1},\Psi_{J2}]=
\int d^3r \left[\frac{(\vec{\nabla} \times \vec{\alpha}_{q1})^2}{2t_2\theta^2} +\frac{(\vec{\nabla} \times \vec{\alpha}_{q2})^2}{2t_1\theta^2} -\frac{i }{\theta}\vec{\alpha}_{q1}\cdot (\vec{\nabla} \times \vec{\alpha}_{q2})\right]\\
&&~~~~~~~~~~~~~~~~ +\int d^3r \left[\gamma_1|(\vec{\nabla}-i\vec{\alpha}_{q1})\Psi_{J1}|^2 + \gamma_2|(\vec{\nabla}-i\vec{\alpha}_{q2})\Psi_{J2}|^2+m_1|\Psi_{J1}|^2+m_2|\Psi_{J2}|^2+({\rm quartic~terms})\right],\nonumber
\end{eqnarray}
\end{widetext}
where we used continuum complex-valued fields $\Psi_{J1}$, $\Psi_{J2}$ to represent the matter that was represented on the lattice by the integer-valued currents $\vec{J}_1$, $\vec{J}_2$, and we did not write the quartic terms explicitly. This is the action for two gauge fields with mutual Chern-Simons interactions, and two matter fields, minimally coupled to the gauge fields.  Here $\gamma_1$, $\gamma_2$, $m_1$, $m_2$ are some effective parameters; along the self-dual line we have $\gamma_1=\gamma_2$ and $m_1=m_2$. For gapped $\Psi_{J1}$, $\Psi_{J2}$, we can integrate these out and obtain a long-wavelength description of (0) in terms of the $\vec{\alpha}_q$ variables. Condensation of $\Psi_{J1}$, $\Psi_{J2}$ leads to phase (IV). We therefore propose Eq.~(\ref{CSQft}) as the field theory describing the transition at the lower multicritical point. As discussed in Sec. \ref{sec:results}, our results on the nature of the (IV)-(0) transition are still conflicting, but we hope that they will stimulate further numerical and analytical\cite{Zhang1989, WenWu1993, ChenFisherWu1993, Sachdev1998,Barkeshli2012} studies. 

\subsection{Phase (III) and the (III)$\rightarrow$(IV) transition}
\label{subsec:IV-III}
To get another perspective on phase (III), we first interpret the coefficient on the last term of Eq.~(\ref{Sdd}) as a statistical interaction for the $\vec{Q}$ variables, given by
\begin{equation}
\theta_{\rm dual}(k)=\frac{-(2\pi)^2\theta(k)}{|\vec{f}_k|^2v_1(k)v_2(k)+\theta^2(k)}
\label{tdual}
\end{equation}
We can shift this coefficient by $2\pi n$, for integer $n$, without changing the Boltzmann weight $e^{-S}$.
This gives us the following equation for the new statistical angle:
\begin{equation}
\theta_{\rm dual,shifted}(k)=\frac{(2\pi)^2|\vec{f}_k|^2v_1(k)v_2(k)}{\theta\left[|\vec{f}_k|^2v_1(k)v_2(k)+\theta^2\right]},
\label{tq}
\end{equation}
where we have used $\theta=2\pi/n$. Performing formal duality on Eq.~(\ref{Sdd}) using the new statistical interaction $\theta_{\rm dual, shifted}$ gives precisely Eq.~(\ref{Smm}).
[The non-commutation of the duality and shift of $\theta$ by multiple of $2\pi$ is well known in the literature\cite{CardyRabinovici1982, Cardy1982, Shapere1989, Fradkin_SL2Z, Witten2003} and is known to correspond to possibility of ``oblique confinement''.  Here we explicitly see this relation by identifying precise bindings of objects in phase (III) as described in Sec.~\ref{sec:reformulations}.]
This allows us to interpret the variables $\vec{M}$ that are gapped in phase (III) as being dual to the $\vec{Q}$ variables after the shift. The precise meaning of the duality is as in the Appendix, and can be viewed as replacing the integer-valued $\vec{Q}$ variables by real-valued variables $\vec{q}$, while maintaining the information about integer-valuedness in terms of new integer-valued $\vec{M}$. We can write $\vec{q}=\frac{\vec{\nabla}\times\vec{\beta}_q}{2\pi}$, and obtain the following action:
\begin{widetext}
\begin{eqnarray}
\label{CSQbeta}
&&S^{(\theta=2\pi/n)}[\vec{\beta}_{q1},\vec{\beta}_{q2},\vec{M}_1,\vec{M}_2]=\frac{1}{2}\sum_k \left[\frac{v_2(k)|[\vec{\nabla} \times \vec{\beta}_{q1}](k)|^2}{|\vec{f}_k|^2v_1(k)v_2(k)+\theta^2} +\frac{v_1(k)|[\vec{\nabla} \times \vec{\beta}_{q2}](k)|^2}{|\vec{f}_k|^2v_1(k)v_2(k)+\theta^2}\right] \\
&&~~~~~+i\sum_k \frac{v_1(k)v_2(k)}{\theta\left[|\vec{f}_k|^2v_1(k)v_2(k)+\theta^2\right]} [\vec{\nabla}\times\vec{\nabla} \times \vec{\beta}_{q1}](-k) \cdot [\vec{\nabla}\times\vec{\beta}_{q2}](k)
+ i\sum_k \left[\vec{M}_1(-k)\cdot\vec{\beta}_{q1}(k) + \vec{M}_2(-k)\cdot\vec{\beta}_{q2}(k)\right].\nonumber
\end{eqnarray}
\end{widetext}
Compared to Eq.~(\ref{CSQ}), we have used different labels for the gauge fields even though the first terms are the same, to emphasize that the coarse-graining procedure will have a different meaning after the shift in $\theta_{\rm dual}$.
Again, we can cast this action into a more familiar form by returning to real space while taking the long-wavelength limit and replacing the current-loop representation with complex matter fields $\Psi_{M1}$ and $\Psi_{M2}$, minimally coupled to $\vec{\beta}_{q1}$ and $\vec{\beta}_{q2}$:
\begin{widetext}
\begin{eqnarray}
\label{CSQftbeta}
&&S^{(\theta=2\pi/n)}_{\rm eff}[\vec{\beta}_{q1},\vec{\beta}_{q2},\Psi_{M1},\Psi_{M2}]= 
\int d^3r \left[\frac{(\vec{\nabla} \times \vec{\beta}_{q1})^2}{2t_2\theta^2} +\frac{(\vec{\nabla} \times \vec{\beta}_{q2})^2}{2t_1\theta^2} 
+\frac{i}{\theta^3 t_1t_2} (\vec{\nabla}\times\vec{\beta}_{q1})\cdot (\vec{\nabla}\times\vec{\nabla} \times \vec{\beta}_{q2})\right] \\
&&~~~~~~~~~~~~~~~~ + \int d^3r \left[ \tilde{\gamma}_1|(\vec{\nabla}-i\vec{\beta}_{q1})\Psi_{M1}|^2 +\tilde{\gamma}_2|(\vec{\nabla}-i\vec{\beta}_{q2})\Psi_{M2}|^2+\tilde{m}_1|\Psi_{M1}|^2+\tilde{m}_2|\Psi_{M2}|^2 +({\rm quartic~terms})\right]. \nonumber
\end{eqnarray}
\end{widetext}

In phase (III) the $\Psi_{M1}$, $\Psi_{M2}$ fields are gapped and we can integrate them out to obtain a description in terms of two gauge fields $\vec{\beta}_{q1}$ and $\vec{\beta}_{q2}$, with a higher-order mutual Chern-Simons term.\cite{Alicea2005, Alicea2006} The latter is irrelevant compared to the Maxwell terms and does not gap the gauge fields. We thus have two gapless modes. 
The transition from phase (III) to phase (IV) is a condensation of the $\Psi_{M1}$ and $\Psi_{M2}$ variables, and along the self-dual line the two fields condense simultaneously. 
We conjecture that the higher-order mutual Chern-Simons term is irrelevant at this transition as well, and therefore the transition is two decoupled inverted XY transitions.\cite{Dasgupta1981, HLM, Radzihovsky1995} 
Returning to $\vec{Q}_1$, $\vec{Q}_2$ variables, we conjecture that the transition from phase (IV) to phase (III) is two decoupled XY transitions where the mutual statistical interaction of $\vec{Q}_1$ and $\vec{Q}_2$, given by $\theta_{\rm dual,~shifted}$ in Eq.~(\ref{tq}) is irrelevant at long wavelengths.  
This interpretation, together with Eq.~(\ref{Q2J}), explains why we are able to use finite-size arguments on the data in Fig.~\ref{R42} to study the properties of the (IV)-(III) phase transition and conclude that it is continuous. If this interpretation is correct, it also means that the phases (IV), (III), (I), and (II) all meet at a single multicritical point.

We remark that while the lattice actions Eqs.~(\ref{CSQ}) and (\ref{CSQbeta}) are mathematically equivalent and contain all phases in Fig.~\ref{phase}, the continuum field theories Eqs.~(\ref{CSQft}) and (\ref{CSQftbeta}) are distinct and apply only near the corresponding multi-critical points.

\section{Irreducible Responses}
\label{sec:irred}
The current-current correlators $C^{ab}_{\mu\nu}$ represent the response of the current $J_{a\mu}$ to an externally applied field $A^{\rm ext}_{b\nu}$. In systems with long-range interactions it is useful to study ``irreducible responses'' $C^{ab,{\rm irred}}_{\mu\nu}$, which are the responses of the currents to the total field $A^{\rm tot}$, made up of both $A^{\rm ext}$ and an internal gauge field induced by the other currents in the system.\cite{MurthyShankarRMP, Herzog2007, Ranged_Loops}  In our model, the statistical interaction is the long-range interaction, and it acts between different loop species in perpendicular current directions. In this section we will derive the appropriate expressions for the irreducible responses, and show their behavior in our system.

If we apply external fields coupled to both species of loops, as in Eq.~(\ref{Aext}), then by the Kubo formula the response of the current variables is given by:
\begin{equation}
\langle J_{a\mu}(k)\rangle=-i\sum_{b,\nu}C^{ab}_{\mu\nu}(k)A^{\rm ext}_{b\nu}(k).
\label{kubo}
\end{equation}
For concreteness, we will assume that $k$ is in the $z$ direction, $k=(0, 0, k_z)$, and this implies that $C^{ab}_{\mu\nu}=0$ if $\mu$ or $\nu$ are in the $z$ direction. As discussed in Sec.~\ref{sec:model}, the lattice mirror symmetries of our action mean that the only correlators which are non-zero in Eq.~(\ref{kubo}) are $C^{aa}_{\mu\mu}$ and $C^{12}_{\mu\nu}$ with $\mu\neq\nu$. This implies that for a gauge field in one direction, we need only to consider its effects on two of the six possible $J_{a\mu}$; for concreteness in this work we will consider $J_{1x}$ and $J_{2y}$. This allows us to write Eq.~(\ref{kubo}) in the following way:
\begin{eqnarray}
\label{matrixdef}
&& \langle \mathbf J\rangle=-i\mathbf{CA}^{\rm ext}, \\
 \langle \mathbf J\rangle\equiv\left [ \begin{array}{c} \langle J_{1x}\rangle \\ \langle J_{2y}\rangle \end{array}\right], 
\mathbf C&\equiv&\left[ \begin{array}{cc} C^{11}_{xx} & C^{12}_{xy}\\ -C^{12}_{xy} & C^{22}_{yy} \end{array}\right],
\mathbf A^{\rm ext}\equiv\left[ \begin{array}{c} A^{\rm ext}_{1x} \\ A^{\rm ext}_{2y} \end{array} \right], \nonumber
\end{eqnarray} 
where we have used the fact that $C^{21}_{yx}=-C^{12}_{xy}$, which we can also deduce from the mirror symmetries of our model. To characterize the response of $\langle \mathbf J\rangle$ to the total field, we write
\begin{equation}
\langle \mathbf J\rangle=-i\mathbf{C}^{\rm irred}\mathbf{A}^{\rm tot},
\label{Cirreddef}
\end{equation}
with $\mathbf{C}^{\rm irred}$ and $\mathbf{A}^{\rm tot}$ defined similarly to the quantities in Eq.~(\ref{matrixdef}). Here $A^{\rm tot}_{a\mu}$ is the total field, and is identified as 
\begin{equation}
A^{\rm tot}_{a\mu}=A^{\rm ext}_{a\mu}+\langle \alpha_{qa\mu} \rangle,
\label{alpha}
\end{equation}
where the gauge fields $\alpha_{qa\mu}$ are precisely those in Eq.~(\ref{CSQ}) mediating the $\vec{J}_1$ and $\vec{J}_2$ interactions. We can calculate the expectation values $\langle \vec{\alpha}_{q}\rangle$ in the presence of $\vec{A}^{\rm ext}$ by analyzing Gaussian integrals in Eq.~(\ref{CSQ}); thus, for any fixed $\vec{J}$ we obtain the following:
\begin{eqnarray}
&&\la \mbox{\boldmath$\alpha$} \ra = -i\mathbf {VJ}\\
\la \mbox{\boldmath$\alpha$} \ra &\equiv&\left [ \begin{array}{c} \la \alpha_{1x} \ra \\ \la \alpha_{2y} \ra \end{array}\right], \quad
\mathbf V\equiv \left[ \begin{array}{cc} v_1(k) & \frac{\theta(k)}{2\sin(k_z/2)} \\ \frac{-\theta(k)}{2\sin(k_z/2)} & v_2(k) \end{array} \right].\nonumber
\end{eqnarray}
Inserting this into Eq.~(\ref{alpha}) and using Eq.~(\ref{matrixdef}) we get
\begin{equation}
\mathbf A^{\rm tot}=(\mathbbm{1}-\mathbf{VC})\mathbf A^{\rm ext},
\end{equation}
and comparing Eqs.~(\ref{matrixdef}) and (\ref{Cirreddef})
gives our final expression for the irreducible responses:
\begin{equation}
\mathbf{C}^{\rm irred}=\mathbf{C}(\mathbbm{1}-\mathbf{VC})^{-1}.
\end{equation}

\begin{figure}[t]
\includegraphics[angle=-90,width=\linewidth]{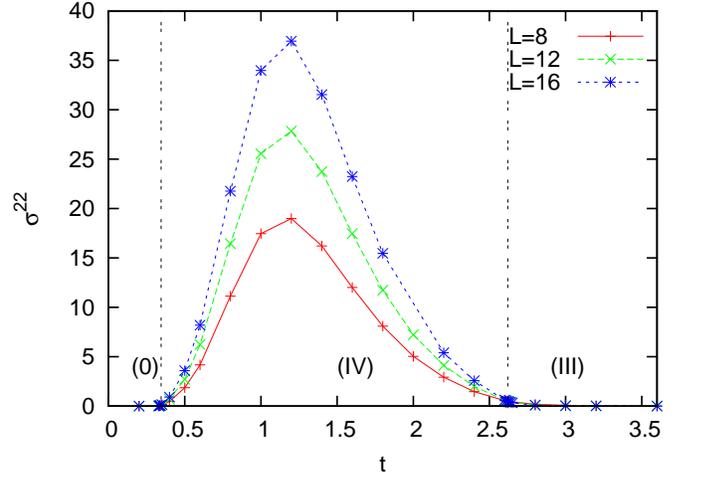}
\caption{
The conductivity $\sigma^{22}$, along the self-dual line, obtained from the raw data in Figs.~\ref{C22} and ~\ref{fig:C12}. Vertical lines mark the phase boundaries. We can see that $\sigma^{22}$ diverges in phase (IV), but is zero in phase (III).}
\label{xxsigma}
\end{figure}

\begin{figure}[t]
\includegraphics[angle=-90,width=\linewidth]{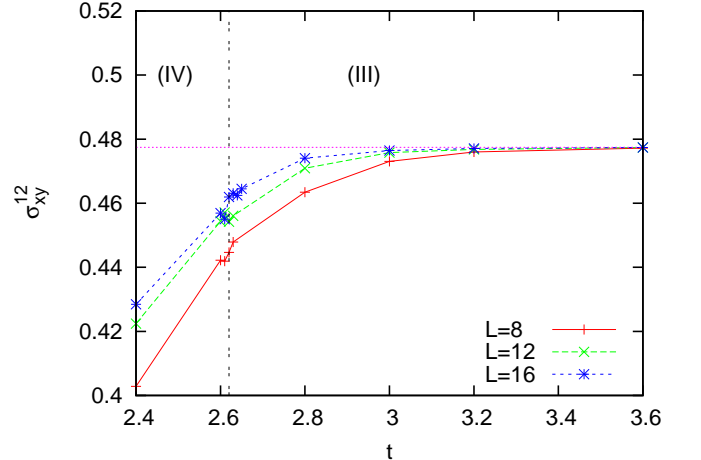}
\caption{The transverse conductivity $\sigma^{12}_{xy}$, along the self-dual line, at the boundary between phases (IV) and (III). A vertical line marks the phase boundary. In phase (IV) $\sigma^{12}_{xy}$ approaches a non-universal value, while in phase (III) it approaches the universal value $3/2\pi$, which is indicated by the horizontal line in the figure. Calculations of $\sigma^{12}_{xy}$ in phase (IV) involve cancellations of similar quantities, which greatly increases the noise in this region. The origin of the cancellation is in the quantization of the $C^{12}$ values in this phase, as seen in Fig.~\ref{C12}. Therefore we have chosen not to display our data in phase (IV).}
\label{xysigma}
\end{figure}

We can use the irreducible responses to determine the conductivities of the system, through the relation
\begin{equation}
\mbox{\boldmath$\sigma$}=\frac{\mathbf{C}^{\rm irred}}{|\vec{f}_k|}.
\end{equation}
We have plotted the diagonal and off-diagonal conductivities along the self-dual line in Figs.~\ref{xxsigma} and \ref{xysigma}.
As in Ref.~\onlinecite{Ranged_Loops}, we can use these conductivities to detect condensation in systems with long-range interactions. We can see that $\sigma^{22}$ diverges with the system size and thus detects the condensation of $\vec{J}_1$, $\vec{J}_2$ in phase (IV), while we recall from Fig.~\ref{C22} that the correlator $C^{22}$ did not. The diagonal conductivity has a crossing at $t_1=t_2\approx0.338$, which is tentatively the location of the transition between phases (0) and (IV). 

It is interesting to note that in phase (III), $\sigma^{22}=0$ while $\sigma^{12}_{xy}$ approaches a universal value of $1/\theta=3/(2\pi)$. We can loosely interpret this if we recall that phase (III) is a condensate of composite objects containing $n$ particles of one type bound to one antivortex of the other type [see Fig.~\ref{bound} and Eq.~(\ref{M2})]. For example, consider a situation where we have a $\vec{J}_1$ charge current flowing in the $x$-direction. This current can be carried by the condensate of the bound states, in which case there is also a $\vec{Q}_2$ current in the $x$-direction, given by $\langle Q_{2x}\rangle=-\langle J_{1x}\rangle/n$. In the absence of any other currents, we get $A_{1x}^{\rm tot}=0$. Furthermore, we can think of the $\vec{Q}_2$ variables as magnetic fluxes for the $\vec{J}_2$ charges. Therefore, by Faraday's law there is an electric field (acting on the $\vec{J}_2$ charges) induced perpendicular to the direction of $\vec{Q}_2$, and we get $-i k_z A_{2y}^{\rm tot} = -2\pi \langle Q_{2x}\rangle = \frac{2\pi}{n}J_{1x}$. This is exactly what we would expect from the conductivity that we derived, $\mathbf A^{\rm tot}=i(\mathbf C^{\rm irred})^{-1} \langle \mathbf J \rangle$.

We can consider the responses for the dual $\vec{Q}_1$, $\vec{Q}_2$ variables as well.  Focusing on a pair $Q_{1y}$, $Q_{2x}$, we define
\begin{eqnarray}
\mathbf C_{\rm dual}\equiv\left[\begin{array}{cc}
\langle Q_{1y}(k)Q_{1y}(-k)\rangle & \langle Q_{1y}(k)Q_{2x}(-k)\rangle \\
-\langle Q_{1y}(k)Q_{2x}(-k)\rangle & \langle Q_{2x}(k)Q_{2x}(-k)\rangle \end{array} \right] ~, \nonumber
\end{eqnarray}
where we used $\langle Q_{2x}(k) Q_{1y}(-k)\rangle = -\langle Q_{1y}(k) Q_{2x}(-k)\rangle$.  The interaction matrix for the specific ordering of the cartesian components is
\begin{eqnarray}
\mathbf V_{\rm dual} \equiv 
\left[ \begin{array}{cc} 
v_{1, {\rm dual}}(k) & \frac{-\theta_{\rm dual}(k)}{2\sin(k_z/2)} \\
\frac{\theta_{\rm dual}(k)}{2\sin(k_z/2)} & v_{2, {\rm dual}}(k) 
\end{array} \right]
= \frac{(2\pi)^2}{|\vec{f}_k|^2} ({\mathbf V}^{-1})^T,
\end{eqnarray}
where the last relation was derived by using Eqs.~(\ref{vdual}) and (\ref{tdual}), and the superscript ``$T$'' denotes the matrix transpose.  The dual and direct responses satisfy the relation
\begin{eqnarray}
\label{VCeq}
\mathbf{VC} + \mathbf{C}_{\rm dual}^T \mathbf{V}_{\rm dual}^T = \mathbbm{1},
\end{eqnarray}
which we can check by using Eqs.~(\ref{QtoJ2}) and (\ref{C12inQ}).  Relation~(\ref{VCeq}) is similar to the relation satisfied in the one-component case.\cite{HoveSudbo2000, Herzog2007, Ranged_Loops}  We can also verify that the irreducible conductivities satisfy 
\begin{equation}
\mbox{\boldmath$\sigma$}\mbox{\boldmath$\sigma$}_{\rm dual}^T=\frac{\mathbbm{1}}{(2\pi)^2},
\end{equation}
which is similar to the relation that conductivities obey when there is only one species of loop.\cite{MurthyShankarRMP, Herzog2007, Ranged_Loops}

\section{Discussion}
\label{sec:concl}
It is instructive to compare these results to those of our earlier study at $\theta=\pi$.\cite{Geraedts} The Boltzmann weight $e^{-S}$ of the $\theta=\pi$ model is invariant under $(\vec{J}_1, \vec{J}_2) \to (-\vec{J}_1, \vec{J}_2)$, while this is not satisfied in the present model. We can see that the correlation between different currents, $C^{12}$, changes sign under this operation, and therefore must be zero in the $\theta=\pi$ model. This explains why that model did not contain the phase (IV) that we have seen in the present study. The location of phase (III) in the two models is also quantitatively different, since the loops must condense in different strengths to avoid the statistical interaction, and this happens at different values of $t$. Phase (III) itself is qualitatively similar in the two models (except for the charge multiplicity in the bound states), and is detected by $C^{12, {\rm irred}}$ in the $\theta=\pi$ model despite the fact that $C^{12}$ is strictly zero.  

From these studies, we can anticipate the behavior of the model with short-range interactions at general statistical angle $\theta\neq\pi$. We expect that the phase diagram will be similar to the one in Fig.~\ref{phase}, except that the phase (III) will feature condensation of more complex composites\cite{CardyRabinovici1982, Cardy1982,Shapere1989} and will occur at different values of $t$. An open question in the present work is the nature of the lower multi-critical point, where our results are conflicting between first and second-order scenarios.  It would be interesting to explore this phase transition in short-ranged models for other values of $\theta$ numerically and analytically.

It is also interesting to explore behavior for more general interactions, particularly for self-dual models with $\vec{J}_1 \leftrightarrow \vec{J}_2$ interchange symmetry. For the model with short-range interactions, we have seen that the statistical interaction qualitatively changes the nature of the phases and phase transitions. On the other hand, for loops with long-ranged interactions decaying as $1/r$ in real space (behaving as $1/k^2$ for small $k$ in momentum space), we expect that the statistical interactions are less important, since here the density fluctuations are very strongly suppressed and the mutual statistics phases are fluctuating less.\cite{Alicea2005, Alicea2006}  In fact, starting with the original model Eq.~(\ref{action}) with short-ranged interactions at $\theta=2\pi/n$, our reformulation in terms of $\vec{M}_1$ and $\vec{M}_2$ variables in Eq.~(\ref{Smm}) can be viewed as precisely such a new model with long-ranged interactions and $\theta_{M1,M2}=2\pi/n$, so the present numerical study already provides information about such a model with $\theta_{M1,M2}=2\pi/3$.  In the absence of the statistical interactions, loops with long-ranged interactions would condense via independent one-component Higgs transitions (inverted XY transitions).\cite{Dasgupta1981, HLM, Radzihovsky1995}  From our discussion in Sec.~\ref{sec:cft}, we conjecture that this remains true also in the presence of the statistical interactions with $\theta \neq \pi$, i.e., they are irrelevant at the phase transition in the long-ranged case.

An interesting case is obtained for marginally long-ranged interactions decaying as $1/r^2$ in real space (behaving as $1/|k|$ for small $k$ in momentum space).\cite{Fradkin_SL2Z,Kuklov2005}  In a recent paper~[\onlinecite{Ranged_Loops}], we studied condensation of single species with such marginal interactions and found second-order transitions with continuously varying critical properties that depend on the coupling of the long-range interaction.  We would like to study condensation for two species with mutual statistics and ask whether the transitions remain continuous for $\theta\neq 0$ and explore the critical properties (which will likely vary with $\theta$).  We can construct a lattice model where we know the phase boundaries exactly from duality considerations\cite{CardyRabinovici1982, Cardy1982, Shapere1989} and can focus on such studies precisely at the transitions.  An interesting question is what happens for $\theta=\pi$ in such models with marginally long-ranged interactions, whether we find a critical loop state or phase separation.  The latter happened in a specific model with short-ranged interactions that we studied in Ref.~\onlinecite{Geraedts}, while we would like to explore if a critical state can be obtained for modified interactions.

For broader outlook, our system is an example where certain reformulations allow direct study of particles with mutual statistics.  It would be interesting to look for other cases where such reformulations may be possible.  Systems with more complex anyons could be interesting,~\cite{Wen2000, Levin2005, Burnell2011, Barkeshli2010, Nayak2008_rmp, Gils2009}, and such combined numerical and analytical studies could bring insights about broader phase diagrams and phase transitions involving topological phases.
Furthermore, the present two-loop system can be viewed as an example of more general actions with topological terms.  In fact, as discussed in Ref.~\onlinecite{Senthil2006_theta}, the two-loop model with $\theta = \pi$ statistical interaction is equivalent to an anisotropic O(4) sigma model with a topological $\theta=\pi$ term; our loop models can be viewed as providing precise lattice realization of this topological field theory\cite{Geraedts, Xu2011} and show that it is important to examine different phases such a theory may have.  Inspired by our two-loop systems, it would be interesting to study precise lattice (discretized space-time) formulations of other topological field theories of current interest\cite{Ng1994, Hansson2004, Cho2011, Chen2011, Kou2009, Wen2000, Burnell2011, Barkeshli2010, LevinStern2009, Neupert2011, ChoDyon} also in other space-time dimensionalities, and ask if they may also allow sign-free reformulations and hence unbiased numerical studies.


\acknowledgments
We are thankful to A.~Vishwanath, M.~P.~A.~Fisher, T.~Senthil, R.~Kaul, G.~Murthy, J.~Moore, N.~Read, A.~Shapere, and W.~Witzak-Krempa for stimulating discussions. This research is supported by the National Science Foundation through grant DMR-0907145; by the Caltech Institute of Quantum Information and Matter, an NSF Physics Frontiers Center with support of the Gordon and Betty Moore Foundation; and by the XSEDE computational initiative grant TG-DMR110052.

\appendix
\section{Formal duality procedure}
\label{app:duality}
This appendix summarizes our duality procedure for one loop species.\cite{PolyakovBook, Peskin1978, Dasgupta1981, FisherLee1989, LeeFisher1989, artphoton}  The original degrees of freedom are conserved integer-valued currents $\vec{J}(r)$ residing on links of a simple 3D cubic lattice; $\vec{\nabla} \cdot \vec{J}(r) = 0$ for any $r$.  To be precise, we use periodic boundary conditions and also require vanishing total current, $\vec{J}_{\rm tot} \equiv \sum_r \vec{J}(r) = 0$.  We define duality mapping as an exact rewriting of the partition sum in terms of new integer-valued currents $\vec{Q}(R)$ residing on links of a dual lattice and also satisfying $\vec{\nabla} \cdot \vec{Q}(R) = 0$ for any $R$ and $\vec{Q}_{\rm tot} = 0$:
\begin{eqnarray}
Z &\!=\!& \sum_{\vec{J}}^\prime e^{-S_{\rm orig}[\vec{J}]} = \sum_{\vec{Q}}^\prime e^{-S_{\rm dual}[\vec{Q}]} , ~~ \label{ZJZQ} \\
e^{-S_{\rm dual}[\vec{Q} = \vec{\nabla} \times \vec{p}]} &\!=\!& \int_{-\infty}^\infty [{\cal D} \vec{j}]^\prime e^{-S_{\rm orig}[\vec{j}]} e^{-i \sum_r \vec{j}(r) \cdot 2\pi \vec{p}(r)} . ~~~~~~ \label{Dj}
\label{dualitystep}
\end{eqnarray}
In the first line, the primes on the sums signify the above constraints on the currents $\vec{J}$ and $\vec{Q}$ respectively.  In the second line, the prime on the real-valued integration measure signifies corresponding linear constraints realized with Dirac delta functions, $\Pi_{r\neq 0} \delta[\vec{\nabla} \cdot \vec{j}(r) = 0]$ and $\delta(\vec{j}_{\rm tot} = 0)$.  For any configuration $\vec{Q}$ satisfying the above constraints, we can find $\vec{p}(r)$ such that $\vec{Q} = \vec{\nabla} \times \vec{p}$, and the constraints on $\vec{j}$ guarantee that the right-hand-side of the last equation does not depend on the choice of $\vec{p}$.

Equations~(\ref{ZJZQ})-(\ref{Dj}) provide a precise way to go from integer-valued sums with {\it constrained} $\vec{J}$ to real-valued integrals with constrained $\vec{j}$, which is achieved with the help of new integer-valued constrained fields $\vec{Q}$.  A formal demonstration can be sketched, e.g., as follows:  We first implement the constraints on $\vec{J}$ using conjugate $2\pi$-periodic phase variables.  We then replace sums over integer-valued $J_\mu(r)$ with integrals over real-valued $j_\mu(r)$ containing a factor $\sum_{p_\mu(r) = -\infty}^\infty e^{-i j_\mu(r) 2\pi p_\mu(r)}$ for each link.  We group configurations $\vec{p}(r)$ into classes specified by $\vec{Q} = \vec{\nabla} \times \vec{p}$ and use summation over members in each class to effectively extend the integrations over phase variables to the full real line.  The latter integrals finally lead to the delta function constraints on the real-valued fields $\vec{j}$ defining the measure $[{\cal D} \vec{j}]^\prime$.  In the process, we see that $\vec{Q}$ can be interpreted as vortex lines in the phase variables conjugate to $\vec{J}$.

An immediate important application is to the case with
\begin{equation}
S_{\rm orig}[\vec{J}] = \frac{1}{2} \sum_k v(k) |\vec{J}(k)|^2 +
i\sum_k \vec{J}(-k)\cdot \vec{A}^{\rm ext}(k) ~,
\label{SJ}
\end{equation}
where we have also coupled the original currents to an external probe gauge field $\vec{A}^{\rm ext}$.  The integration over $\vec{j}$ in Eq.~(\ref{Dj}) is Gaussian and readily gives basic averages
\begin{equation}
\la j_\mu(k) j_{\mu'}(k') \ra_0 = \frac{\delta_{k+k'=0}}{v(k)} \left(\delta_{\mu \mu'} - \frac{f_{k,\mu} f_{k,\mu'}^*}{|\vec{f}_k|^2} \right) ~,
\end{equation}
where $f_{k,\mu} \equiv 1 - e^{i k_\mu}$.  We then obtain
\begin{equation}
S_{\rm dual}[\vec{Q}] = \frac{1}{2} \sum_k \frac{[2\pi\vec{Q}(-k) + \vec{B}(-k)] \cdot [2\pi\vec{Q}(k) + \vec{B}(k)]}{v(k) |\vec{f}_k|^2} ,
\label{SQ}
\end{equation}
where $\vec{B} \equiv \vec{\nabla} \times \vec{A}^{\rm ext}$.  The relation between Eq.~(\ref{SJ}) and Eq.~(\ref{SQ}) is what we call ``duality map'' in the main text.

\bibliography{bib4twoloops}
\end{document}